\documentclass[preprint,12pt]{elsarticle}

\usepackage{amssymb}

\usepackage{mathpazo} 
\usepackage{amsmath}
\usepackage[ruled,vlined]{algorithm2e}

\journal{Nuclear Physics A}

\begin{document}

\begin{frontmatter}



\title{Improved methodologies to study the performance of the ANET Compact Neutron Collimator}


\author[inst1,inst2]{O. Sans-Planell}
\author[inst3,inst4]{F. Cantini}
\author[inst1,inst2]{M. Costa}
\author[inst1,inst2]{E. Durisi}
\author[inst3,inst4]{F. Grazzi}
\author[inst1,inst2]{E. Mafucci}
\author[inst1,inst2]{V. Monti}
\author[inst5]{R. Bedogni}
\author[inst6]{Y. Li}
\author[inst6]{L. van Eijck}

\affiliation[inst1]{organization={Universita degli Studi di Torino},
            addressline={Via Pietro Giuria 1}, 
            city={Torino},
            postcode={10125}, 
            state={},
            country={Italy}}

\affiliation[inst2]{organization={INFN Sezione di Torino},
            addressline={Via Pietro Giuria 1}, 
            city={Torino},
            postcode={10125}, 
            state={},
            country={Italy}}
            
\affiliation[inst3]{organization={INFN Frascati National Laboratories},
            addressline={Via Enrico Fermi 54}, 
            city={Frascati},
            postcode={00044}, 
            state={},
            country={Italy}}
            
\affiliation[inst4]{organization={Consiglio Nazionale delle Ricerche, Istituto dei Sistemi Complessi},
            addressline={Via Madonna del Piano 10}, 
            city={Sesto Fiorentino},
            postcode={50019}, 
            state={},
            country={Italy}}
            
\affiliation[inst5]{organization={INFN Sezione di Firenze},
            addressline={Via Madonna del Piano 10}, 
            city={Sesto Fiorentino},
            postcode={50019}, 
            state={},
            country={Italy}}
            
\affiliation[inst6]{organization={Delft University of Technology},
            addressline={Mekelweg 15}, 
            city={Delft},
            postcode={2629 JB}, 
            state={},
            country={Netherlands}}
            
\affiliation[inst7]{organization={INFN Sezione di Pavia},
            addressline={Via Agostino Bassi 6}, 
            city={Pavia},
            postcode={27100}, 
            state={},
            country={Italy}}
\begin{abstract}
The ANET project aims at developing 2D compact neutron collimators for neutron imaging applications. The results of the ANET collimator performances, presented in this communication, are based on data collected at the FISH beamline at TU-Delft. Two independent methods to evaluate the neutron radiography resolution are described and discussed, as well as a comparison of the beam divergence with or without the ANET collimator.
 
\end{abstract}


\begin{highlights}
\item Neutron Beam Divergence measurement 
\item Numerical method to derive neutron image resolution with a Siemens Star
\item New data-driven approach to defining the MTF threshold on neutron radiography measurements
\end{highlights}

\end{frontmatter}
\section{Introduction}
The ANET project aims at developing 2D Compact Neutron Collimators (CNC) for Neutron Imaging applications. Its design and operational principles have been described in \cite{Bedogni2021design}, while the first results on its performance have been reported in \cite{sans2021first}.\\
This paper illustrates two improved methodologies to determine the image resolution from measured data. A PSI Siemens-Star and a Gd knife-edge assembly are used as reference samples \cite{siemens-star}. The applicability of the herein exposed methods is general and not restricted to a specific beam-line or collimation system. Data have been collected at the FISH beamline \cite{zhou2018fish} at the HOR research reactor, owned by Technische Universiteit Delft. In the following sections, after having illustrated the experimental setup, a detailed description of the analysis methods is given. The ANET CNC effectiveness, in terms of beam divergence reduction and consequent resolution improvement, is also demonstrated.

\section{The experimental setup}
\begin{figure}[h]
    \centering
    \includegraphics[width=1\textwidth]{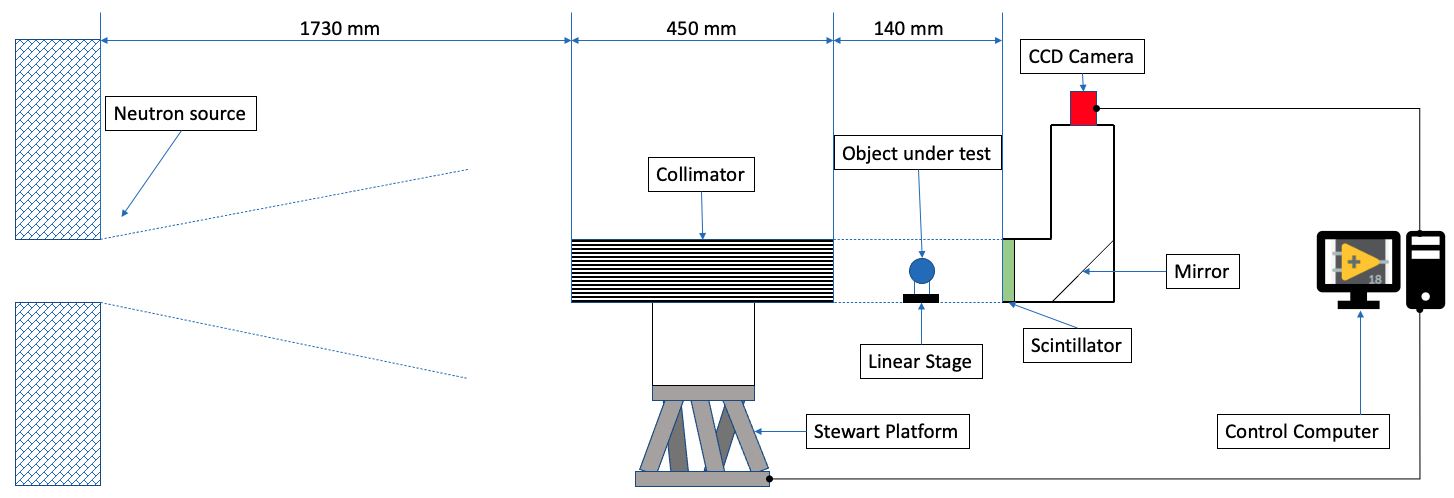}
    \caption{Scheme of the set-up used for the measurements (not in scale). The ANET collimator is mounted on top of a Stewart Platform and raised at the beam axis height by a calibrated spacer. The samples are installed on top of X-Y remotely controlled moving stages} 
    \label{fig:Set-up-design}
\end{figure}

The FISH beam-line provides, at the standard detector position (5.5 $m$ from the shutter), a thermal neutron flux of 3$\cdot$10$^6cm^{-2}s^{-1}$, with a rectangular field of view of 50 x 100 $mm^2$, resulting in an asymmetric divergence for the horizontal and vertical axis: the quoted divergence angles are 0.176 and 0.207 degrees respectively.\\
This is a very interesting working case for the application of the ANET CNC due to its intrinsic 2D symmetric geometry. For this measuring campaign, a neutron imaging system with a 100$\mu m$ thick LiF/ZnS(Ag) scintillator and a 16-bit ANDOR camera had been installed at 232$cm$ from the beam shutter, as shown in figure \ref{fig:Set-up-design}. The distance of the ANET CNC from the shutter has been constrained by hardware limitations on the beam-line. To measure the horizontal and vertical divergence of the neutron beam,  with and without the ANET CNC, the procedure explained in \cite{kaestner2017samples} has been applied. It requires the acquisition of neutron radiographies of a reference sample at several distances from the detector, in an appropriate range, as specified in the following.
\\The ANET CNC has been  operated following the procedure detailed in \cite{sans2021first}, in which a dynamic pattern is used. The exposure time for each radiography has been tuned to  30 seconds without the collimator and 300 seconds with the ANET CNC. The acquisition time has been set to reach approximately one-half of the full dynamic range of the ANDOR camera. For each reference sample, eight different radiographs  were taken, from 10$mm$ to 105$mm$ away from the scintillator. To evaluate the impact of the ANET CNC in terms of resolution and beam divergence, two  independent analysis methods have been applied. They are described in the following sections.
\section{The "unfolded" Siemens Star method}
\begin{figure}[h]
    \centering
    \includegraphics[width=0.7\textwidth]{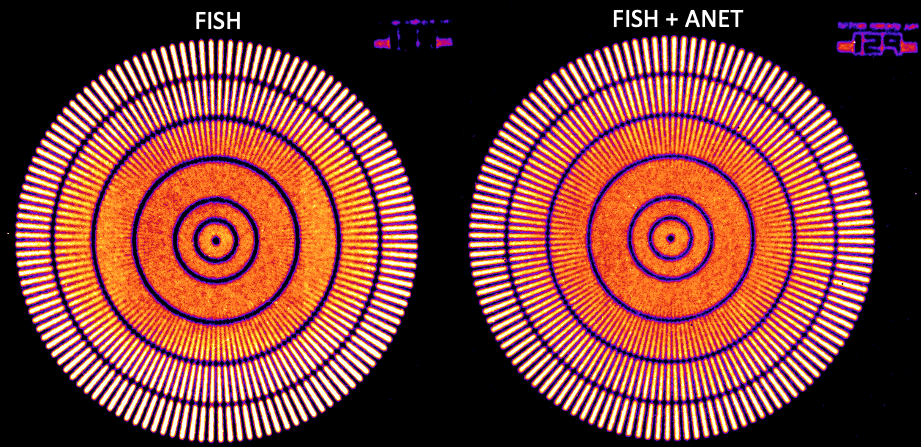}    \caption{Comparison between two Siemens star measurements at FISH, without (left) and with (right) the ANET CNC. The distance of the sample  from the detector is at 20$mm$.} 
    \label{fig:Siemens-comparison}
\end{figure}

\begin{figure}[h]
    \centering
    \includegraphics[width=0.8\textwidth]{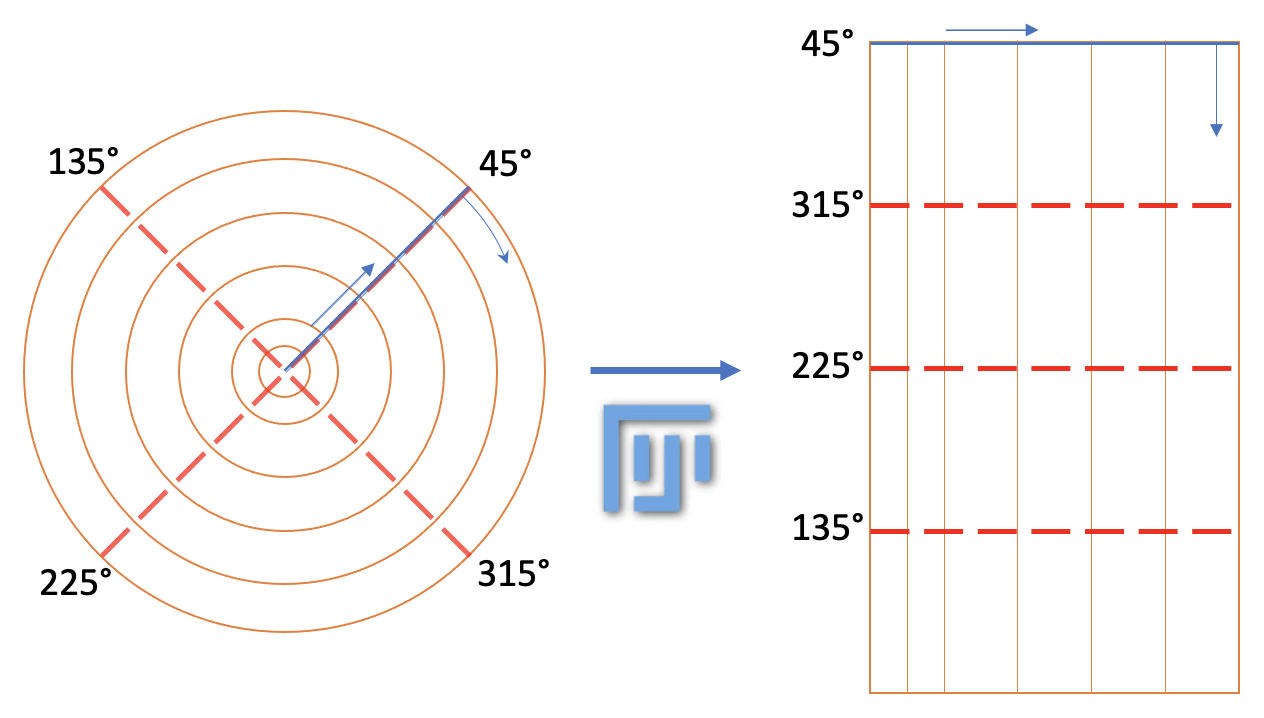}
    \caption{Representation of the process of "unfolding" the Siemens star using FIJI. The reslicing is done clockwise from the 45$^{\circ}$ line.} 
    \label{fig:Siemens-unfold}
\end{figure}

\begin{figure}[h]
    \centering
    \includegraphics[width=1\textwidth]{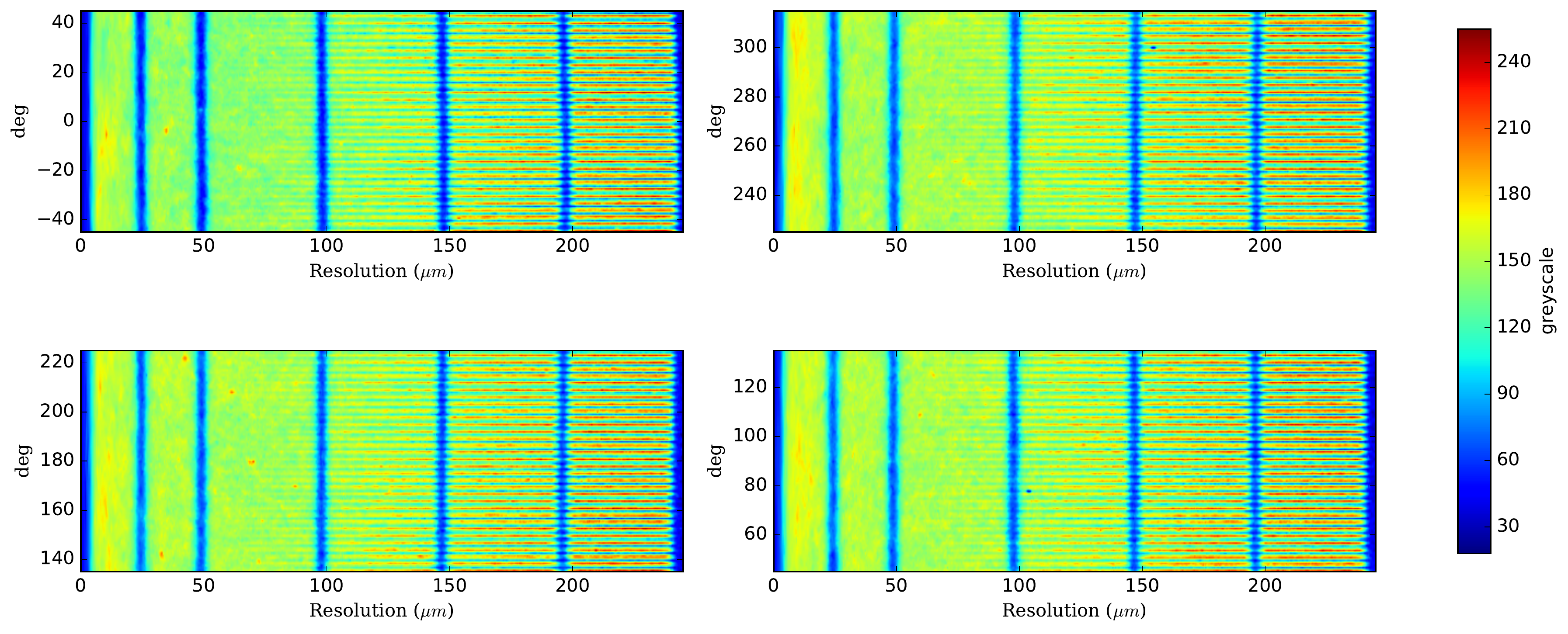}
    \caption{Visualisation in four angular slices of the "unfolded" Siemens star. The greyscale image is displayed using the "jet" LUT in order to highlight the difference between the noise region and the actual spikes of the Siemens star.} 
    \label{fig:Reslice}
\end{figure}

\begin{figure}[h]
    \centering
    \includegraphics[width=1\textwidth]{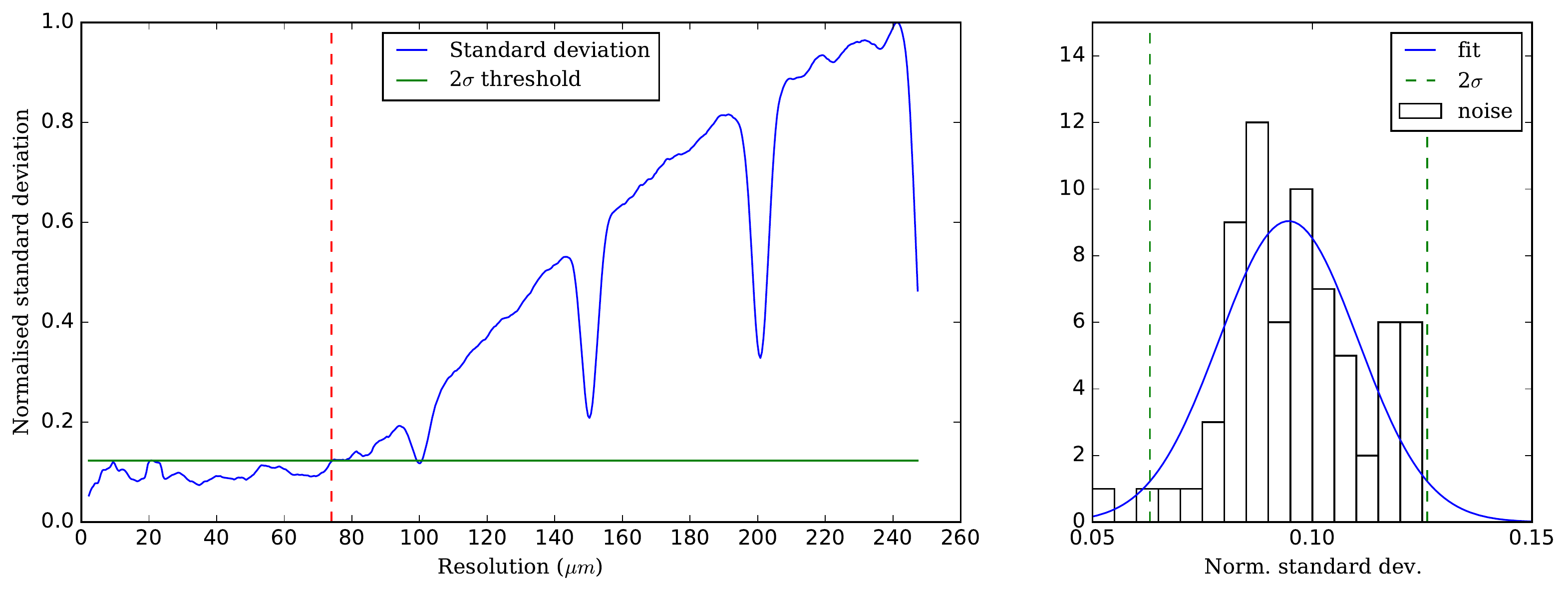}
    \caption{Left: Sample of the standard deviation of a measure at 10$mm$ from the detector, without the ANET CNC present. The red dotted line separates the noise region and the signal. Using a 2$\sigma$ threshold, the value for the resolution is 76$\mu m$. Right: Histogram of the noise created by the standard deviation in the noise region. The noise is fitted with a Gaussian curve in order to obtain the 2$\sigma$ threshold.} 
    \label{fig:Siemens-std}
\end{figure}

In figure \ref{fig:Siemens-comparison} two radiographs are shown, which show "by-eye" the improvement in resolution produced when the ANET CNC is placed in front of the sample. 
By choosing the appropriate ImageJ Look-Up-Table (LUT) it is possible to highlight the blurry areas and separate them from the well-defined regions of the Siemens star. The right image in which ANET is included in the set-up shows an evident improvement in spatial resolution with respect to the image on the left, with FISH alone.\\
The precision to appreciate the resolution on the Siemens star by optical means is limited to $\pm25\mu m$. Further, it's not always easy to distinguish at which point the threshold is, especially when considering the difference between the vertical and horizontal divergences in a beam-line with a rectangular shutter or pin-hole such as that present at FISH. In order to improve the evaluation of spatial resolution, a new numerical approach has been taken.\\
By means of the software FIJI ~\cite{imagej}, the single image has been radially re-sliced from the centre to the most external circle, starting at 45$^{\circ}$ angle. The procedure is represented in figure \ref{fig:Siemens-unfold}. 
The image contains, on the right, a representation of the Siemens star "unfolded", where the top pixel row is equivalent to the 45$^{\circ}$ radius used to start the re-slicement. Dividing the image in four 90$^{\circ}$  slices allows to separate the two components of the divergence. Figure \ref{fig:Reslice} shows the 4 images representing the resolution in the 4 different sectors of the Siemens star.
\newline To improve the best-possible optical precision of $\pm$25$\mu m$, the following method is adopted: for each single image, the standard deviation of every vertical array of points is taken (from left to right). The standard deviation is a measure of the amount of dispersion of a set of values, and thus the expectation will be constant and small on the first points, in the inner area of the Siemens star (inner rings, leftmost area in each figure), where only noise is present. The standard deviation value will increase with the higher contrast (outer rings, right-most area in each figure). This is shown in figure \ref{fig:Siemens-std}, where the Standard deviation is plotted against the resolution.
In the figure, the initial region is almost constant and it is possible to distinguish the valleys made by the Siemens star ring markers. The decrease in the standard deviation at those points is due to the fact that the points in the area corresponding to the incisions are more uniform than that of the surroundings, thus decreasing the variation. The quoted value for the spatial resolution is found at the intersection between the constant (on average) area and the sloped one.\newline 
To determine the value of the resolution, a threshold to distinguish signal from noise has been chosen. What has been noticed from the experimental data is that, in the region where there is only noise, the pixel counting has a relatively small standard deviation (low contrast), while in the signal region, the pixel count distribution has a larger standard deviation (high contrast). In figure \ref{fig:Siemens-std} the green line represents the threshold, tuned to the region where only noise is present, set at 2$\sigma$ from the average value of the noise. The vertical red dotted line separates the noise and signal regions. A systematic study to evaluate the sensitivity of the result with respect to the choice of the threshold has been done, varying from 2$\sigma$ to 5$\sigma$, observing a maximum variation on the final L/D of less than 5\%. This possible systematic error is taken into consideration for the error calculation. For the horizontal resolution calculation, the average of top (45 to 135 deg) and bottom (225 to 315 deg) images are used, while for the vertical resolution the left (135 to 225 deg) and right (-45 to 45 deg) are used. \\
The procedure to calculate the effective L/D factor is done by repeating the measurement and varying the distance of the sample from the scintillator, as detailed in \cite{kaestner2017samples}.\\
Finally the divergence angle is extracted through equation \ref{LD-to-div}. 
\begin{equation}
    \theta_{div} = arctan\big(\frac{1}{L/D}\big)
    \label{LD-to-div}
\end{equation}
The horizontal and vertical divergence angles measured with and without the ANET CNC are reported in table \ref{tab:Siemens divergence} 
\begin{table}[h!]
\centering
\begin{tabular}{||c c c c||} 
 \hline
Horizontal   & Horiz+ANET & Vertical & Vert+ANET \\ [0.5ex] 
 \hline\hline
 $0.215\pm 0.011 $  & $0.144\pm 0.007$ & $0.325\pm 0.022$ & $0.162\pm 0.008$ \\ 
  [1ex] 
 \hline
\end{tabular}
\caption{Horizontal and Vertical divergence angles with and without the ANET CNC, measured using the "unfolded" Siemens method.}
\label{tab:Siemens divergence}
\end{table}

\section{The Gadolinium knife-edge technique}
\begin{figure}[h]
	\centering
	\includegraphics[width=0.7\linewidth]{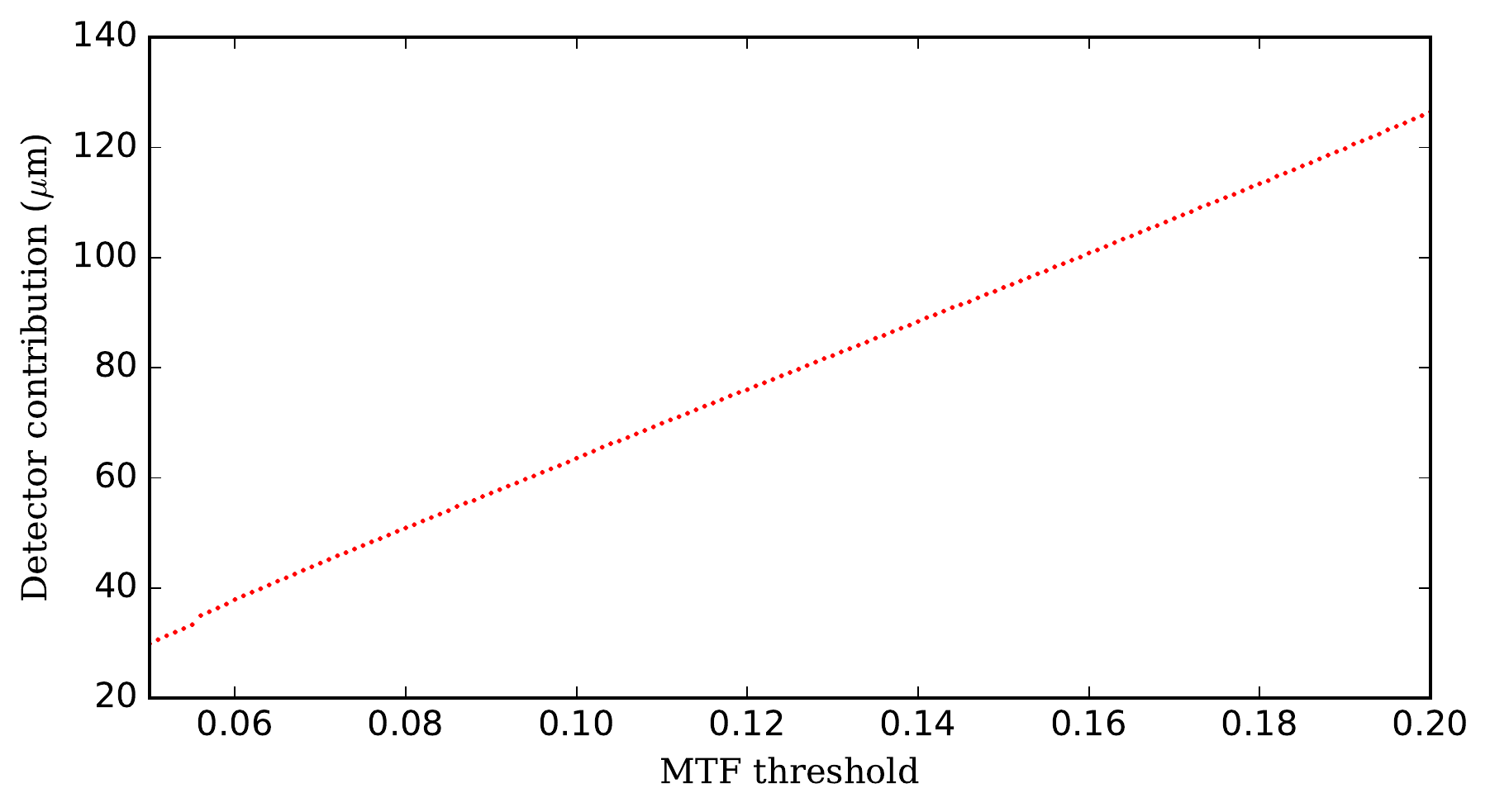}
	\caption{Detector contribution to the spatial resolution measurement as a function of the MTF threshold.}
	\label{fig:MTF_threshold}
\end{figure}

\begin{figure}[h]
	\centering
	\includegraphics[width=0.7\linewidth]{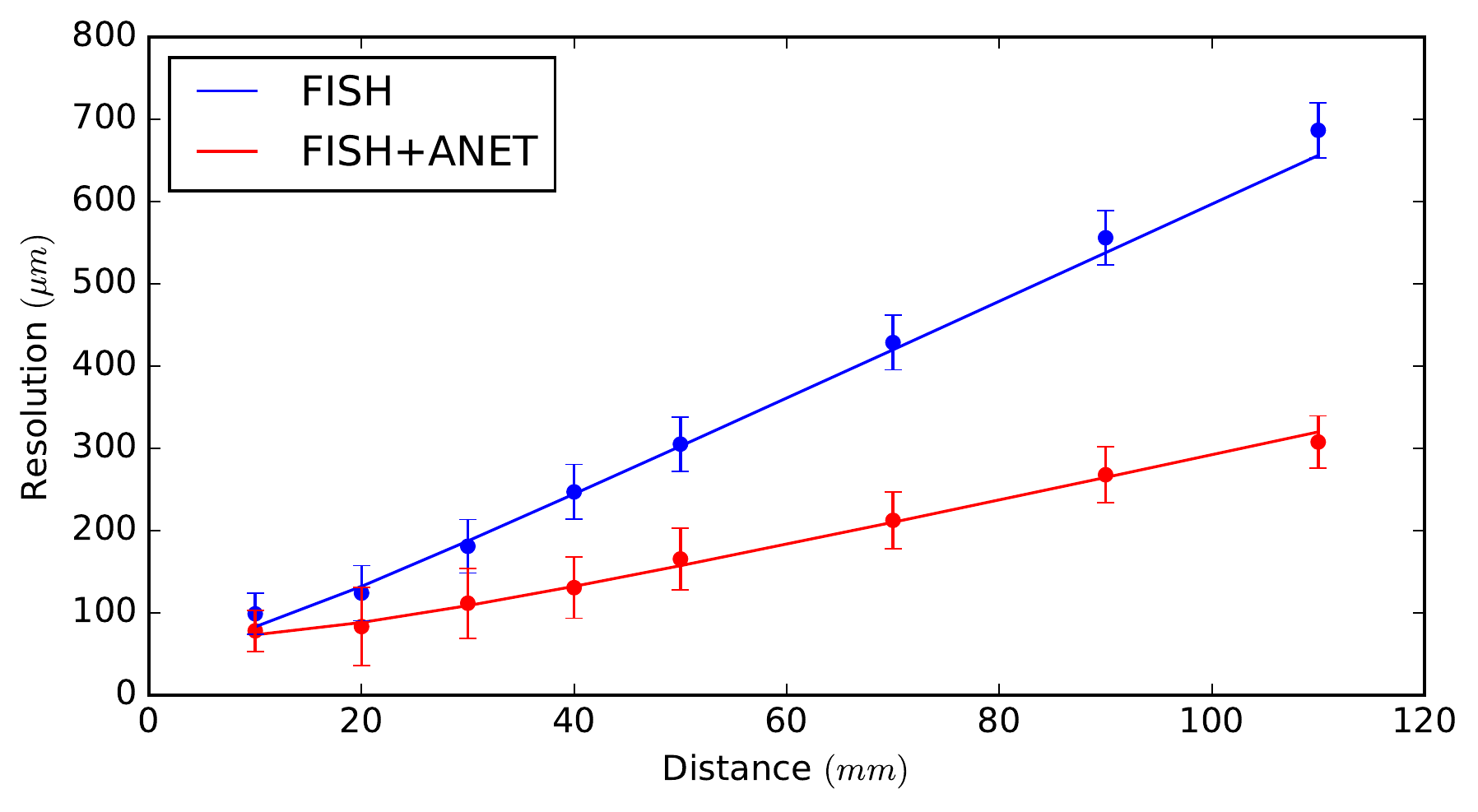}
	\caption{Resolution fit for the vertical measurement of the divergence with and without the ANET CNC.}
	\label{fig:Gd-resolution}
\end{figure}

The gadolinium knife-edge is a reference sample, composed by a 100$\mu m$ thick gadolinium sheet with two polished edges and mounted on an aluminium frame. With a fixed field of view, The edge images from this high contrast edge device  allow a good estimate of the spatial resolution by means of the Modulation Transfer Function (MTF) analysis method\cite{boreman2001modulation}.\\
The measurements, as explained in the previous section, have been performed at 8 different distances from the scintillator, ranging from 10$mm$ up to 105$mm$. \begin{figure}[h]
    \centering
    \includegraphics[width=0.6\textwidth]{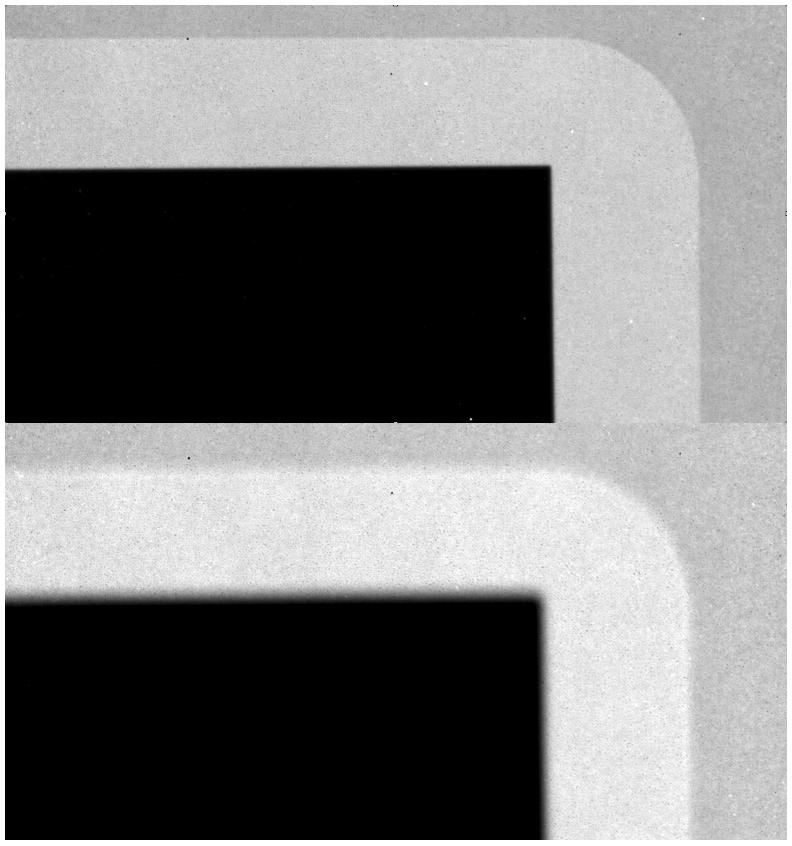}
    \caption{Two examples of radiographs with the ANET collimator, at 10$mm$ (top) and 100$mm$ (bottom) from the detector.}
    \label{fig:Gd-two-examples}
\end{figure}
\newline The gadolinium knife-edge sample has been placed orthogonal to the beam, in order to properly measure the vertical and horizontal spatial resolutions. Figure \ref{fig:Gd-two-examples} shows the variation in image blurring from two measurements (with the ANET CNC) at respective distances from the detector of 10$mm$ (top) and 100$mm$ (bottom).\\
For each distance, two images have been taken, and the MTF has been calculated.\\
The MTF method relies on the setting of a threshold identifiyng the so-called limit resolution, i.e. the resolution at which it becomes impossible to 
separate a line's pair.
Nevertheless, it is well known, there is not a universal standard criteria to set the threshold value. \cite{mtf0,mtf1,mtf2,mtf3}. 
In the following we propose a procedure to determine in an empirical way the proper threshold, relying on an independent measurement of the limit resolution, i.e. the detector blurring. In our case, the previously described measurement with the Siemens star attached to the scintillator delivers a precise value of the limit resolution and it will be used to fix the proper threshold. The procedure has been the following: 
\begin{enumerate}
    \item Set a MTF threshold value in the physical range.
    \item Calculate the MTF for each radiography on the dataset obtained by moving the sample at different distances from the scintillator
    \item Calculate the values for the L/D and the detector contribution using the method described in \cite{kaestner2017samples}
\end{enumerate}
This procedure has been iteratively performed for several threshold values in the range from 0.05 up to 0.25, leading to the graph in figure \ref{fig:MTF_threshold}.
There is a clear dependence between the detector blurring contribution and the choice of the threshold. The measurement with the Siemens star in contact with the scintillator leads to a value for the limit resolution of 50$\mu m$, which in fig \ref{fig:MTF_threshold} corresponds to a threshold value of 0.08.\\ 
Applying this threshold, the graph in figure \ref{fig:Gd-resolution} is generated: a clear improvement in the resolution when including the ANET CNC in the FISH set-up is visible.
\\Consequently, an improvement of the beam divergence, is expected: the horizontal and vertical divergence angles measured with and without the ANET CNC are reported in table ~\ref{tab:Gd divergence}
\begin{table}[h!]
\centering
\begin{tabular}{||c c c c||} 
 \hline
Horizontal   & Horiz+ANET & Vertical & Vert+ANET \\ [0.5ex] 
 \hline\hline
 $0.200\pm 0.008$  & $0.148\pm 0.006$ & $0.354\pm 0.013$ & $0.160\pm 0.006$ \\ 
  [1ex] 
 \hline
\end{tabular}
\caption{Table to test captions and labels.}
\label{tab:Gd divergence}
\end{table}

\section{Results}
\begin{figure}[h]
    \centering
    \includegraphics[width=0.9\textwidth]{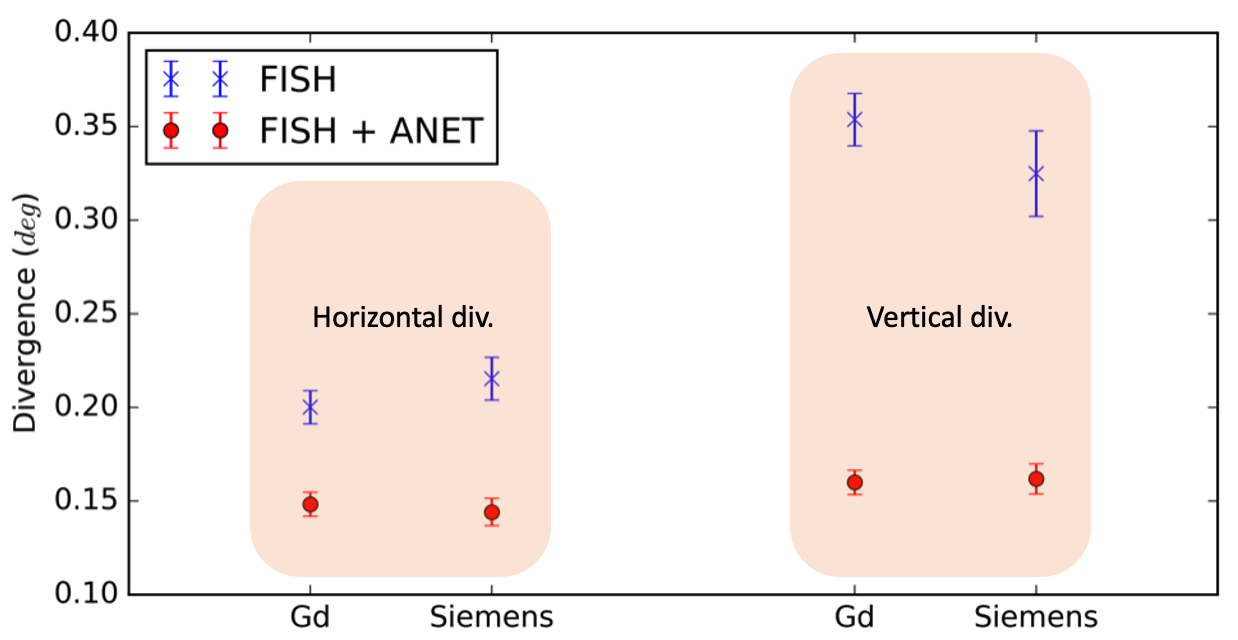}
    \caption{Calculated divergence angle on every configuration using the gadolinium knife-edge method and the Siemens star. The errors have been evaluated through a Monte-Carlo error propagation.}
    \label{fig:LD-Gd}
\end{figure}

The paper presents the performance of the ANET CNC in terms of spatial resolution and beam divergence. Two different reference samples and analysis methods have been used.
The L/D values are extracted for each sample, one for the vertical and one for the horizontal axis. The results are summarized in figure \ref{fig:LD-Gd} where the comparison between the divergences measured at FISH with and without the ANET CNC are shown.
\\The divergence measurements obtained using the Siemens star are compatible with those obtained using the gadolinium knife-edge.\\
The 2D structure of the ANET CNC tends to make uniform the vertical and horizontal divergence.  It is important to remark that, despite the ANET CNC applying the same theoretical correction on both axes, the effect is more relevant on the vertical axis with respect to the horizontal one, being the horizontal one originally more collimated\\
The ANET CNC has proven to be a valid instrument not only to improve, but also to make uniform the divergence of a neutron beam.  In this contribution, the effect on the divergence has been studied, with two novel methods that allow to estimate the resolution quantitatively with improved control of the inherent parameters.

\
 \bibliographystyle{elsarticle-num} 
 \bibliography{elsarticle-template-num}

\begin{thebibliography}{10}
\expandafter\ifx\csname url\endcsname\relax
  \def\url#1{\texttt{#1}}\fi
\expandafter\ifx\csname urlprefix\endcsname\relax\def\urlprefix{URL }\fi
\expandafter\ifx\csname href\endcsname\relax
  \def\href#1#2{#2} \def\path#1{#1}\fi

\bibitem{Bedogni2021design}
R.~Bedogni, M.~Costa, E.~Durisi, F.~Grazzi, A.~Lega, E.~Mafucci, L.~Menzio,
  V.~Monti, O.~S. Planell, L.~Visca, Design of a compact neutron collimator,
  Journal of Instrumentation 16~(08) (2021) P08055.

\bibitem{sans2021first}
O.~Sans-Planell, M.~Costa, E.~Durisi, E.~Mafucci, L.~Menzio, V.~Monti,
  L.~Visca, F.~Grazzi, R.~Bedogni, S.~Altieri, First results with the {ANET}
  compact thermal neutron collimator, Journal of Instrumentation 16~(11) (2021)
  P11025.

\bibitem{siemens-star}
C.~Grünzweig, G.~Frei, E.~Lehmann, G.~Kühne, C.~David, Highly absorbing
  gadolinium test device to characterize the performance of neutron imaging
  detector systems, Review of Scientific Instruments (2007) 78.

\bibitem{zhou2018fish}
Z.~Zhou, J.~Plomp, L.~van Eijck, P.~Vontobel, R.~P. Harti, E.~Lehmann,
  C.~Pappas, Fish: A thermal neutron imaging station at hor delft, Journal of
  Archaeological Science: Reports 20 (2018) 369--373.

\bibitem{kaestner2017samples}
A.~Kaestner, Z.~Kis, M.~Radebe, D.~Mannes, J.~Hovind, C.~Gr{\"u}nzweig,
  N.~Kardjilov, E.~Lehmann, Samples to determine the resolution of neutron
  radiography and tomography, Physics Procedia 88 (2017) 258--265.

\bibitem{imagej}
M.~D. Abr{\`a}moff, P.~J. Magalh{\~a}es, S.~J. Ram, Image processing with
  imagej, Biophotonics international 11~(7) (2004) 36--42.

\bibitem{boreman2001modulation}
G.~D. Boreman, Modulation transfer function in optical and electro-optical
  systems, Vol.~4, SPIE press Bellingham, WA, 2001.

\bibitem{mtf0}
S.~Williams, A.~Hilger, N.~Kardjilov, I.~Manke, M.~Strobl, P.~Douissard,
  T.~Martin, H.~Riesemeier, J.~Banhart, Detection system for microimaging with
  neutrons, Journal of instrumentation 7~(02) (2012) P02014.

\bibitem{mtf1}
A.~Saad, M.~Saab, D.~Gatinel, Repeatability of measurements with a double-pass
  system, Journal of Cataract \& Refractive Surgery 36~(1) (2010) 28--33.

\bibitem{mtf2}
J.~R. Rogers, Three-bar resolution versus mtf: how different can they be
  anyway?, in: An Optical Believe It or Not: Key Lessons Learned, Vol. 7071,
  SPIE, 2008, pp. 38--45.

\bibitem{mtf3}
J.~H. Kinney, M.~Nichols, U.~Bonse, X-ray tomographic microscopy using
  synchrotron radiation, APS Users Organization Steering Committee (1992) 99.

\end{thebibliography}





\end{document}